\documentclass[preprint,authoryear,12pt]{elsarticle}

\usepackage{epsfig}

\usepackage{amssymb}

\usepackage[%ps2pdf,%
a4paper=true,%
breaklinks=true,%
colorlinks=true,%
pdfauthor={de Mello Neto et al.},%
pdftitle={Measurements of cosmic rays at the highest energies with the Pierre Auger Observatory}%
]{hyperref}

\journal{Advances in Space Research}

\begin{document}

%%%%%%%%%%%%%%%%%%%%%%%%%%%%%%%%%%%%%%%%%%%%%%%%%%%%%%%%%%%%%%%%%%%%%%%%%%%%%
%% Frontmatter
\begin{frontmatter}

%% Title, authors and addresses

% Use the tnoteref command within \title and fnref within \author or \address for footnotes;
% use the corref command within \author for corresponding author footnotes;
% use the ead command for the email address,
% and the form \ead[url] for the home page:
% \title{Title\tnoteref{label1}}
% \tnotetext[label1]{}
% \author{Name\corref{cor1}\fnref{label2}}
% \ead{email address}
% \ead[url]{home page}
% \fntext[label2]{}
% \cortext[cor1]{}
% \address{Address\fnref{label3}}
% \fntext[label3]{}

\title{Measurements of cosmic rays at the highest energies with the Pierre Auger Observatory}
 
% Use optional labels to link authors explicitly to addresses:
% \author[label1,label2]{}
% \address[label1]{}
% \address[label2]{}

\author{de Mello Neto, J. R. T.\footnote{\emph{E-mail address} \texttt{jtmn@if.ufrj.br}}, for the Pierre Auger Collaboration\footnote{ Full author list: http://www.auger.org/archive/authors\_2012\_07.html } }
\address{Instituto de F\'isica, Universidade Federal do Rio de Janeiro, Ilha do Fund\~ao\\
Rio de Janeiro, RJ  Brazil \\
Observatorio Pierre Auger, Av. San Mart\'in Norte 304, 5613 Malarg\"ue, Argentina
}
%\cortext[cor]{Corresponding author}
%\fntext[footnote2]{Additional information regarding the corresponding author}
%\ead{jtmn@if.ufrj.br}

% Url can be given like this:
% \ead[url]{http://www.elsevier.com/wps/find/authorsview.authors/latex}

%\fntext[footnote3]{ Full author list: http://www.auger.org/archive/authors\_2012\_07.html }
%\ead{augerspokersperson@fnal.gov}

\begin{abstract}
%% Text of abstract 
This paper summarizes the  status and the recent measurements from the Pierre Auger Observatory. The energy
spectrum is described and its features discussed. We report  searches for anisotropy of cosmic ray arrival directions  on large scales and through correlation with catalogues of celestial objects.  We also present the search for  anisotropies in the data without the use of astronomical catalogues. The first measurement of the   proton-air cross section around  $10^{18}$ eV is discussed. Finally, the mass composition is addressed with measurements of the variation of the
depth of shower maximum with energy and with the muon density at ground. 
 
\end{abstract}

\begin{keyword}
%first keyword \sep second keyword \sep more keywords
 ultra high energy cosmic rays; cosmic ray experiments
% keywords here, in the form: keyword \sep keyword
% PACS codes here, in the form: \PACS code \sep code
\end{keyword}

\end{frontmatter}

\parindent=0.5 cm

%%%%%%%%%%%%%%%%%%%%%%%%%%%%%%%%%%%%%%%%%%%%%%%%%%%%%%%%%%%%%%%%%%%%%%%%%%%%%
%% Main text
\section{Introduction}

Cosmic rays are  energetic particles from space incident on the Earth's atmosphere, e.g. protons, 
heavier nuclei, photons and neutrinos, plus the secondary particles that are generated as they traverse
the atmosphere. The study of cosmic ray physics had a pioneer role in the study of elementary particles 
and their interactions. The nature and origin of the  Ultra High Energy Cosmic Rays (UHECR - here standing for cosmic rays with energy E$\geq 0.1$ EeV, 1~EeV$\equiv 10^{18}$~eV)   
remain puzzles almost fifty years  since cosmic rays with energies of the order of 100 EeV 
were first reported~\citep{Lins63}. The  UHECRs are the subject of extensive studies including measurements of the energy spectrum,  the nature of the primaries and the identification of possible sources. With energies that can reach about two orders of magnitude higher than the ones possible  in the LHC, the study of the UHECRs probes the hadronic interactions in otherwise inaccessible regions of the phase space. 

\section{The Pierre Auger Observatory}

The Pierre Auger Observatory~\citep{observ} is the world's largest UHECR facility, conceived to measure the flux, arrival direction distribution and mass composition of cosmic rays from 0.1 EeV (low energy extensions included 
\citep{heat} \citep{amiga}) 
to the highest energies with high statistical significance. It is a hybrid detector that combines both surface and fluorescence detectors at the same site.  The surface detector (SD) consists of 1660 10m$^2$ $\times$ 1.2\,m   water-cherenkov stations  deployed over 3000
km$^2$ on a 1500 m triangular grid.   An ``infill'' array with a 750 m grid was added to the SD with the purpose of measuring showers of  lower energy. The SD is overlooked by a fluorescence detector (FD) composed of four fluorescence stations, each one with 6 wide angle telescopes, and one additional station with 3 high-elevation telescopes also conceived to measure showers of lower energy.   The surface detector   stations
sample   the electrons, photons and muons   in the shower front at ground level. The fluorescence telescopes  can record   ultraviolet 
light emitted as the shower
crosses the atmosphere, allowing  one  to observe  the longitudinal development of the air shower.
The fluorescence detector operates only on clear, moonless nights, so its duty cycle is about 13\%. On the other hand, the  surface detector array has a duty cycle close to 100\%.

 \section{The cosmic ray energy spectrum}
 
\begin{figure}
%\centering
\hspace {-7mm}
\begin{tabular}{cc}
\epsfig{file=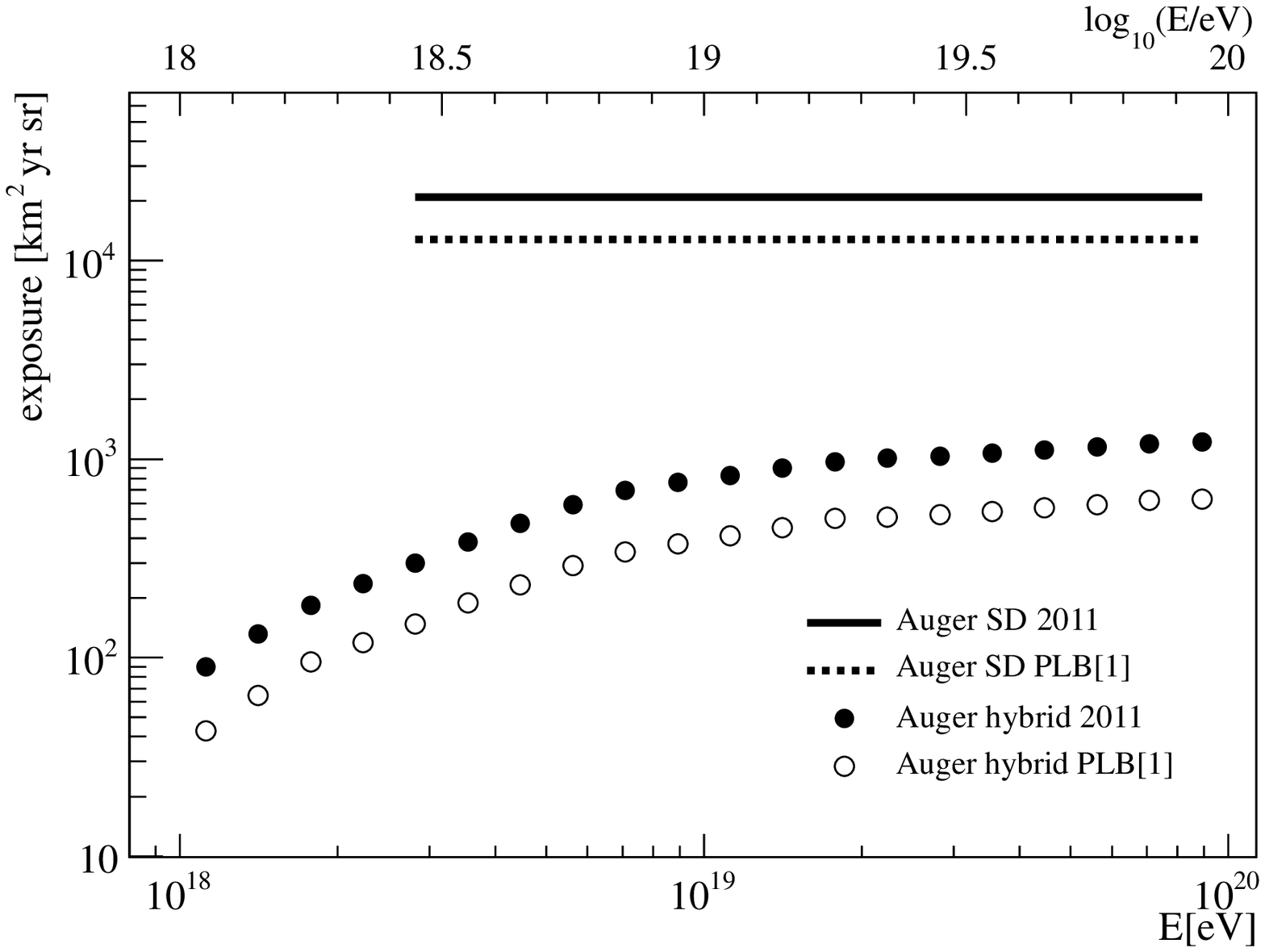,width=0.52\linewidth,clip=} &
\epsfig{file=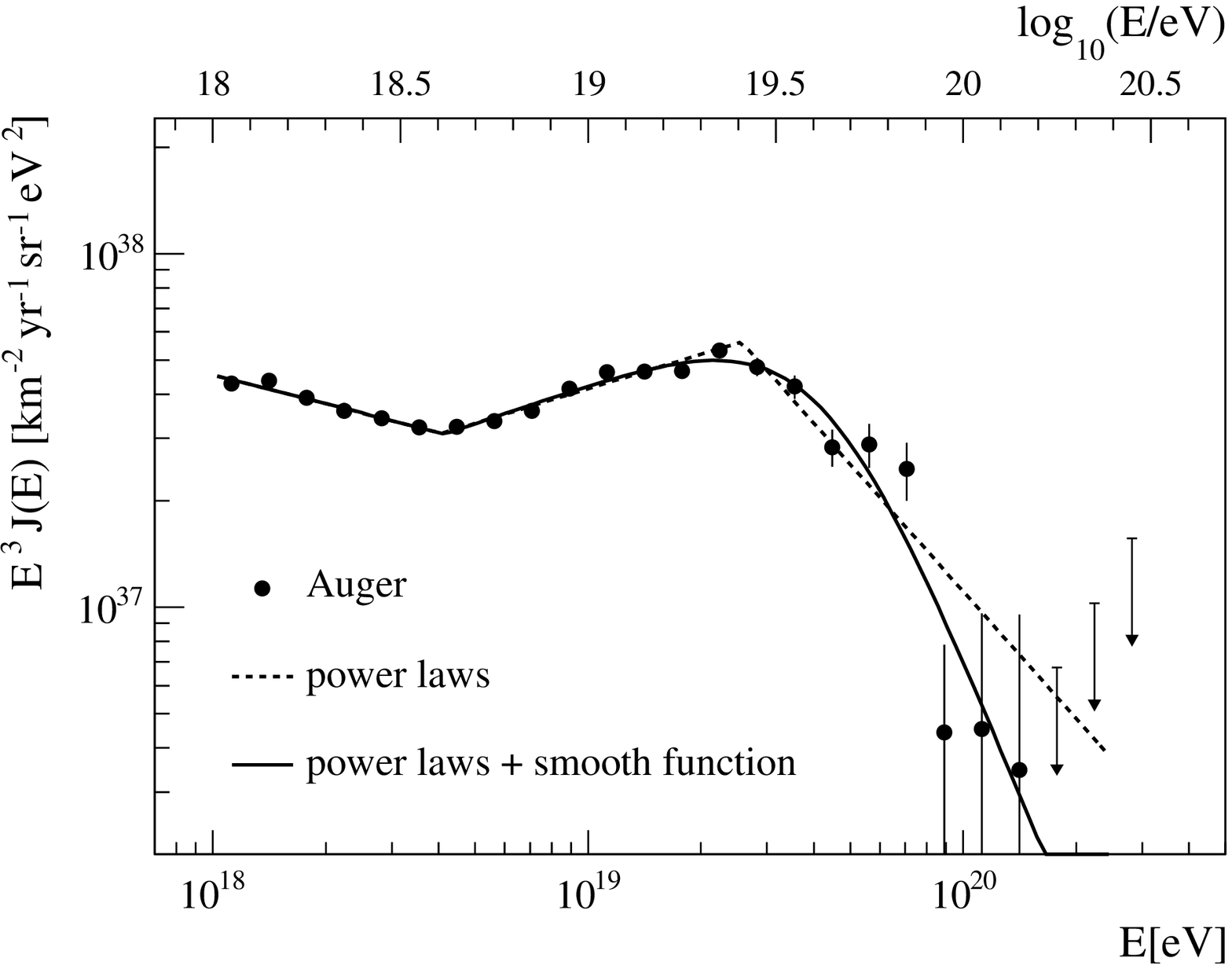,width=0.52\linewidth,clip=} 
\end{tabular}
\caption{ Left: The SD and hybrid exposures used for the current flux measurement compared with a previously published data set \citep{spectrum} . The SD exposure is shown for energies higher than $10^{18.5}$ eV where the detector is fully efficient.  Right: The combined energy spectrum is fitted with two functions. Only statistical uncertainties are shown. The systematic uncertainty in the energy scale is 22\%. \label{f1} }
\end{figure}

It is crucial to measure accurately the cosmic ray flux above $10^{18}$ eV
in order to discriminate among different models describing the transition between galactic and extragalactic cosmic rays and between the suppression induced by the cosmic ray propagation and features of the injection spectrum of the sources. The energy spectrum was measured by the Pierre Auger Observatory and here we report the combined measurements \citep{spectrum}, \citep{salamida}  of the SD and the hybrid data. The SD-only events used in Fig. \ref{f1}  were taken between 1 January 2004 and 31 December 2010, with a total exposure of 20905 km$^2$~sr~yr, while the hybrid events were recorded from 1 November 2005 to 30 September 2010. The higher energies are more easily accessible to the SD that has larger exposure. 
The points at the low end of the plotted spectrum are obtained from the Auger hybrid events.  The hybrid events are defined by the simultaneous triggering of a fluorescence telescope and at least one surface detector station.  The lower energy threshold  and good energy resolution of those events enable measurements to be extended to $10^{18}$ EeV. The exposure of the hybrid mode of the Pierre Auger Observatory has been calculated using a time-dependent Monte Carlo simulation. The changing configurations of both fluorescence and surface detectors are taken into account for the determination of the on-time of the hybrid system, see Fig. \ref{f1} (left plot).
The two spectra were combined using a maximum likelihood method. They have the same systematic uncertainty in the energy scale (22\%), but the normalization uncertainties are independent, 6\% for the SD data and between 6\% and 10\% for the hybrid data. In the region where both distributions overlap (from about $3\times 10^{18}$ eV to about $7\times 10^{19}$ eV) they are mutually consistent within the normalization uncertainties.

A piecewise fit using power laws $(dN/dE \sim E^{-\alpha})$ and another one using power laws plus a smooth function are shown in the plot of  Fig.~\ref{f1} (right plot). We can see that there are two clear spectral features: at 4.1 EeV ($10^{18.61}$ eV)  and 26 EeV ($10^{19.41}~\textrm{eV}$). The first feature is known as the ankle, where the spectral index changes from $\alpha = 3.27 \pm 0.02$ to $2.68 \pm 0.01$.  The spectrum is steeper after the second feature (the power law fit taking the value $\alpha = 4.2 \pm 0.1$) and is compatible with  the "GZK suppression", predicted by Greisen, Zatsepin and Kuz'min \citep{gzk1}, \citep{gzk2} immediately after the discovery of the cosmic microwave background (CMB).  As shown in  Fig.~\ref{f1} (right plot) the combination of the two spectra has enabled the precise measurement of the ankle and of the flux suppression at high energies. 

The physical origin of the ankle is not settled yet. There are two main candidate scenarios to explain it.
In the first scenario  the models relate the ankle to the transition from a galactic component to a harder extragalactic one \citep{linsleyproc}, \citep{hillas}. The second scenario involves  models in which cosmic rays are supposed to be extragalactic protons down to energies below 1 EeV and the concave shape is due to the effect of energy losses of the protons by pair creation with CMB photons, the so-called dip scenario \citep{berez}.   In order to elucidate this issue, anisotropy measurements are very important (and will be discussed in the following session). Measurements of composition are also important because one expects that heavy nuclei dominate galactic cosmic rays at EeV energies, since their confinement by galactic magnetic fields is a rigidity dependent effect.

The suppression at the highest energies, measured with unprecedented statistical significance, 
is consistent with the expectations from the so called "GZK suppression", understood as the attenuation of extragalactic protons by photo-pion production off CMB photons or as the suppression of  nuclei by photodesintegration. However, it must  be noted that the interpretation of the flux suppression in terms of interactions with the CMB does not exclude	additional contributions related to the acceleration mechanism, such as a change in the injection spectrum at the source or the maximum energy of the accelerators. 
 
\section{Studies of  anisotropy in the arrival direction of the UHECRs}

As mentioned in the previous section, if the ankle is due to the change from a galactic to an extragalactic component, a dipolar modulation in the  cosmic ray arrival directions is expected. The amplitude of the dipole should increase with energy up to the ankle, reaching a level of a few percent, even if  primaries are heavy nuclei. Predictions of the amplitude and orientation of the dipole depend on the magnitude of the magnetic field and also on the source distribution. If the dip scenario is the correct explanation for the ankle, it means that UHECRs above $10^{18}$ eV are already dominated by the extragalactic component, and their flux is expected to be highly isotropic. However a dipole of less than one percent amplitude is still expected due to the {\it Compton-Getting effect}. It would produce a dipolar distribution constant with energy and with an orientation related to the relative movement of the Earth with respect to the "rest frame"  of the extragalactic cosmic rays. 

\begin{figure}
%\centering
\hspace {-7mm}
\begin{tabular}{cc}
\epsfig{file=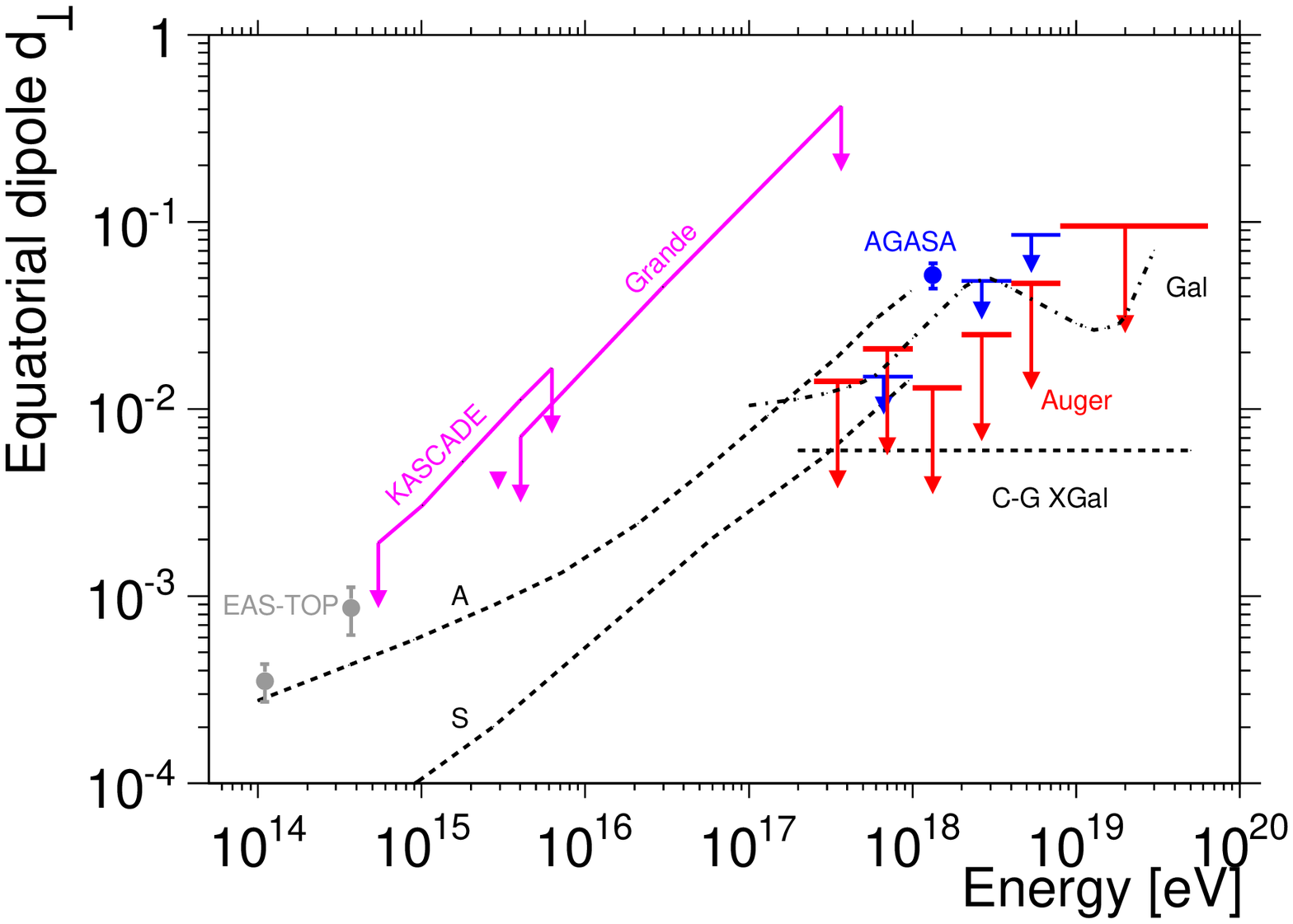,width=0.52\linewidth,clip=} &
\epsfig{file=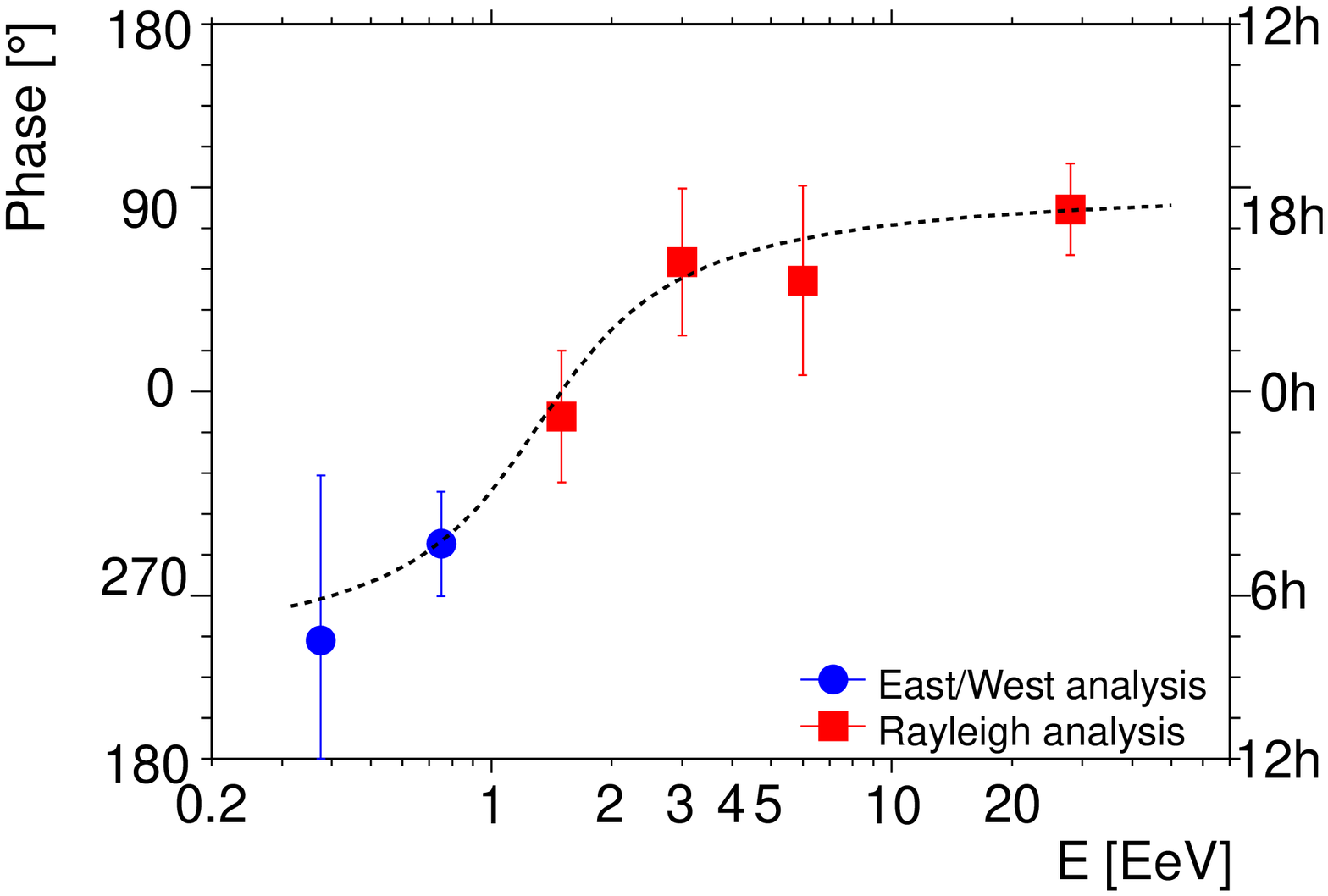,width=0.52\linewidth,clip=} 
\end{tabular}
\caption{ Left: Upper limits on the anisotropy of first harmonic amplitude  as a function of energy and model predictions (see text).    Right: Phase of the first harmonic as a function of energy. The dotted line represents a hypothetical smooth transition of the dipole direction.  \label{f2} }
\end{figure}
The distribution in right ascension of the flux of CRs arriving at a detector can be characterized by the amplitudes and phases of its Fourier expansion. Our aim is to determine the first harmonic amplitude $r$ and its phase $\phi$.  The phase indicates  the direction of the dipole in right ascension. 

Thanks to the high statistics of the SD data a first harmonic analysis was performed in different energy ranges starting from $2.5\times 10^{17}$ eV in search for dipolar modulations in right ascension \citep{dipole}, \citep{lyberis}.  This analysis used the fact that the exposure is almost uniform in right ascension.

Below $E\sim 1$ EeV, the detection efficiency of the array depends on zenith angle and composition, which amplifies detector-dependent variations in the counting rate. Consequently, our results below 1 EeV are derived using simple event counting rate differences between Eastward and Westward directions. That technique using relative rates allows a search for anisotropy in right ascension without requiring any evaluation of the detection efficiency.
 The angular resolution is defined as the angular aperture around the arrival directions of cosmic rays within which 68\% of the showers are reconstructed. At the lowest observed energies, events trigger as few as three surface detectors. The angular resolution of events having such a low multiplicity is contained within 2.2$^\circ$, which is quite sufficient to perform searches for large-scale patterns in arrival directions, and reaches  $\sim$ 1$^\circ$ for events with multiplicities larger than five. 

No significant amplitude was detected. The 99\% CL upper limits as a function of energy are shown in   Fig.~\ref{f2} (left plot) together with predictions   from two different galactic magnetic field models (A and S),  for a purely galactic origin of UHECRs   (Gal), and the expectations from the Compton-Getting effect (C-G Xgal)   \citep{dipole}.  A particular model with an antisymmetric halo magnetic field (A) is already excluded by the upper limits. In  Fig.~\ref{f2} (right  plot)  the phase measurement as a function of the energy shows an interesting pattern: it suggests a smooth transition between a phase of $\sim 270^{\circ}$ (consistent with the right ascension of the galactic center) below $1\times 10^{18}$ eV and another phase  of  $\sim 90^{\circ}$ (consistent with the right ascension of the galactic anti-center) above $5\times 10^{18}$~eV. This is interesting since a real anisotropy would need less events to be established with high statistical confidence through phase consistency in ordered energy intervals rather than  by amplitude measurements  \citep{dipole}. New data will show if this feature still stands.

%\section{Correlation with celestial objects}
\begin{figure}
%\centering
\hspace {-7mm}
\begin{tabular}{cc}
\epsfig{file=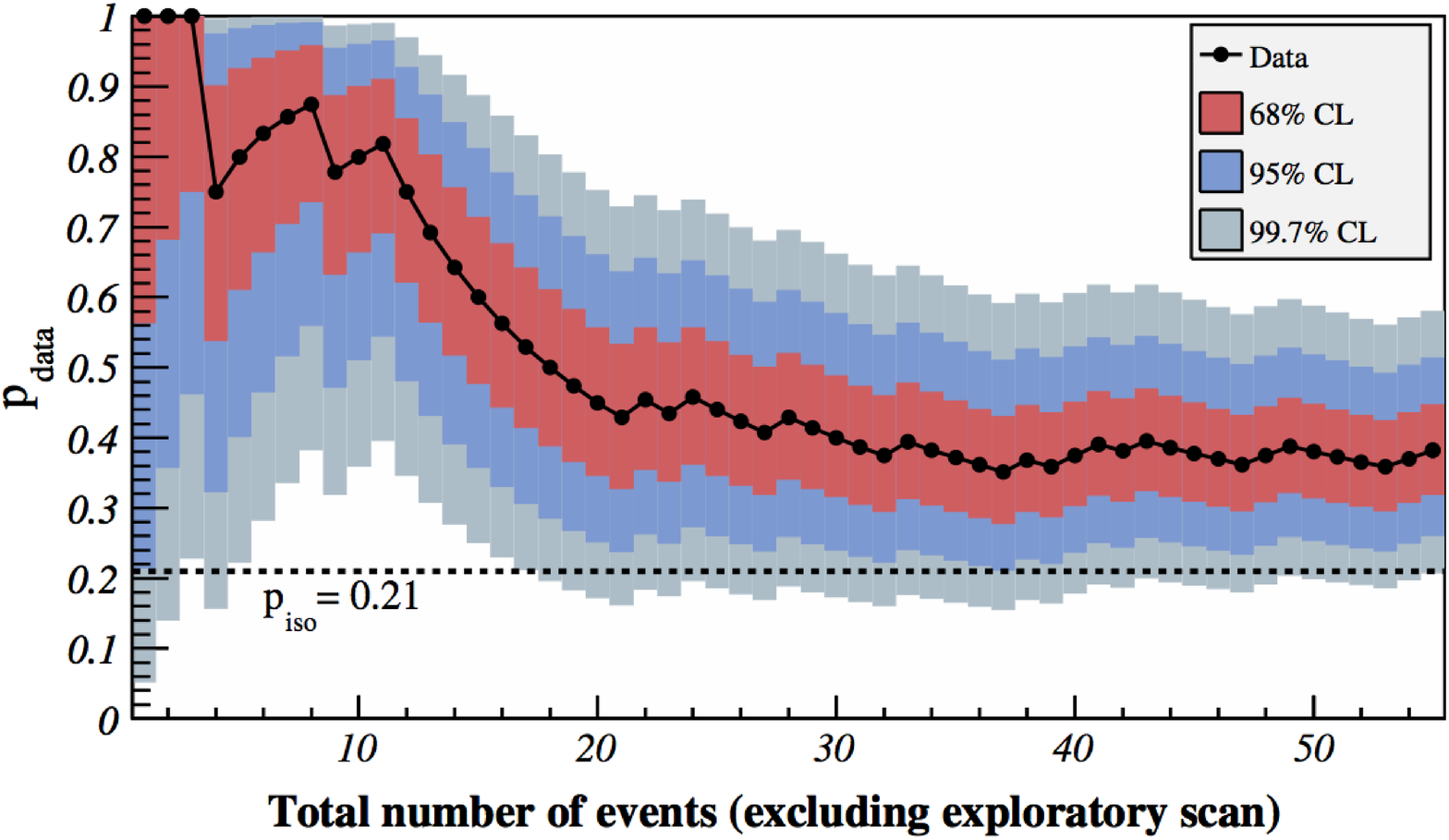,width=0.56 %0.475
\linewidth,clip=} &
\epsfig{file=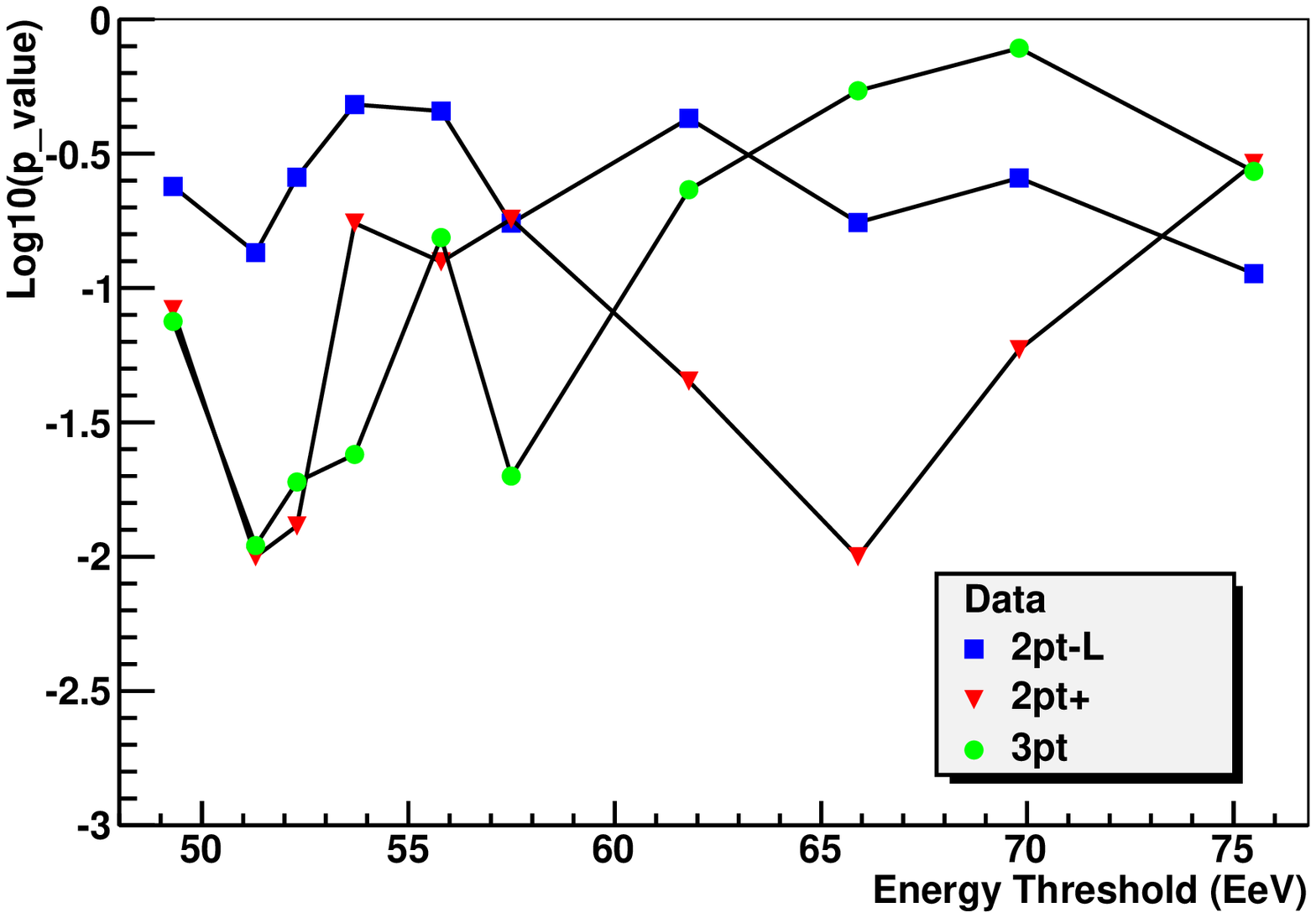,width=0.50 %0.425
\linewidth,clip=} 
\end{tabular}
\caption{ Left:  Fraction of events correlating with AGNs as a function of the cumulative number of events, starting after the exploratory data. The expected correlating fraction for isotropic cosmic rays is shown by the dotted line.  Right:  The minimum in $P_{\mathrm{value}}$ is at 100 events for the 2pt+ and 3pt methods and corresponds to an energy of about $51\approx$ EeV. \label{f3} }
\end{figure}

For energies up to $10^{15}$ eV, cosmic rays are believed to have a galactic origin and shock acceleration in supernova remnants could be the most likely source. At the highest energies, the most probable sources of  ultra-high energy cosmic rays (UHECRs)  are extragalactic:   jets of active galactic nuclei (AGN), galaxy radio lobes, gamma ray bursts and colliding galaxies, among others \citep{kotera}.

 The Pierre Auger Collaboration reported \citep{agn}, \citep{science} a correlation of its highest energy events with the AGNs in the V\'eron-Cetty and V\'eron catalogue. The first 14 events were used for an exploratory scan  that yielded the following search parameters: energy threshold ($E_{th} = 55$ EeV), maximum angular separation ($\Psi \leq 3.1^{\circ}$) and maximum redshift ($z\leq 0.018)$. Those parameters minimize the probability that the correlation with AGN could result from a background fluctuation if the flux were isotropic. The subsequent 13 events established a 99\% confidence level for rejecting the hypothesis of isotropic cosmic ray flux. The reported fraction of correlation events was ($69^{+11}_{-13})$\%. 
By adding data with $E_{th} = 55$ EeV up to the end of 2009 (69 events in total), the  correlation level decreased to $(38^{+7}_{-6})$\%  \citep{update}. For this dataset, we show in  Fig.~\ref{f3} (left plot)   the most likely value of the fraction of the correlated events, plotted with black dots as a function of the total number of time-ordered events (the events used in the exploratory scan are excluded).   The most recent estimate of the fraction of correlating cosmic rays is $(33\pm 5)$\%, with 21\% expected under the isotropic hypothesis \citep{kampertICRC}.  
                                                                                                                                                                                                                                                                                                                                                                                                                                                                                                                                                                                                                                                                                                                            
{\it A posteriori }  studies \citep{update}  showed that the distribution of arrival directions of  the 69 highest energy cosmic rays is compatible with models (for suitable values of two parameters, the smoothing factor $\sigma$ and an isotropic fraction $f_{\textrm{iso}}$) based on  populations of nearby extragalactic objects, such as galaxies in the 2MRS and AGNs in the SWIFT-BAT catalogues. The models fit the data for smoothing angles around a few degrees and for correlating fractions of order 40\% ($f_{\textrm{iso}} \approx 0.6$). 
The data do not fit either the isotropic expectation or the predictions of the models with  $f_{\textrm{iso}}= 0$. 
A large isotropic fraction could indicate that the model is not using a catalogue that includes all the contributing UHECR sources or that a fraction of arrival directions are isotropized by large magnetic deflections of a highly charged particle component.

%\section{Intrinsic anisotropy}
The UHECR anisotropy can also be studied with the use of catalog independent methods. In \citep{intrinsic} three methods, named 2pt-L, 3pt and 2pt+, each giving a different measure of self-clustering in arrival directions, were tested on mock  data sets. The impact of sample size and magnetic smearing was studied. These studies suggested that the three methods could efficiently respond to a real anisotropy in the data with a $P_{\mathrm{value}}$ = 1\% or smaller with a data set of the order of one hundred events.
The methods were applied to the data from the 20 highest energy events to the highest 110. As shown in Fig.~\ref{f3} (right plot),  the two most powerful methods (3pt and 2pt+) show a minimum in the distribution of $P_{\mathrm{value}}$ as a function of the energy threshold for an energy around 51 EeV,  (i.e., at 100 events) \citep{intrinsic}.
 But there is no $P_{\mathrm{value}}$ smaller than 1\% in any of the 30 (correlated) scanned values. There is thus no strong evidence of clustering in the data set which was examined. 
In case there is a true weak  anisotropy in the data, the common minimum   could  indicate  the onset of this anisotropy,  while the less powerful method (2pt-L) would not have  detected it. For higher energy thresholds the number of events decreases and the power of the methods diminish as expected.    Simulations show \citep{intrinsic} that this null result is compatible with the correlation with the VCV catalogue discussed above.

\section{Mass composition}
\begin{figure}
%\centering
\hspace {-7mm}
\begin{tabular}{cc}
\epsfig{file=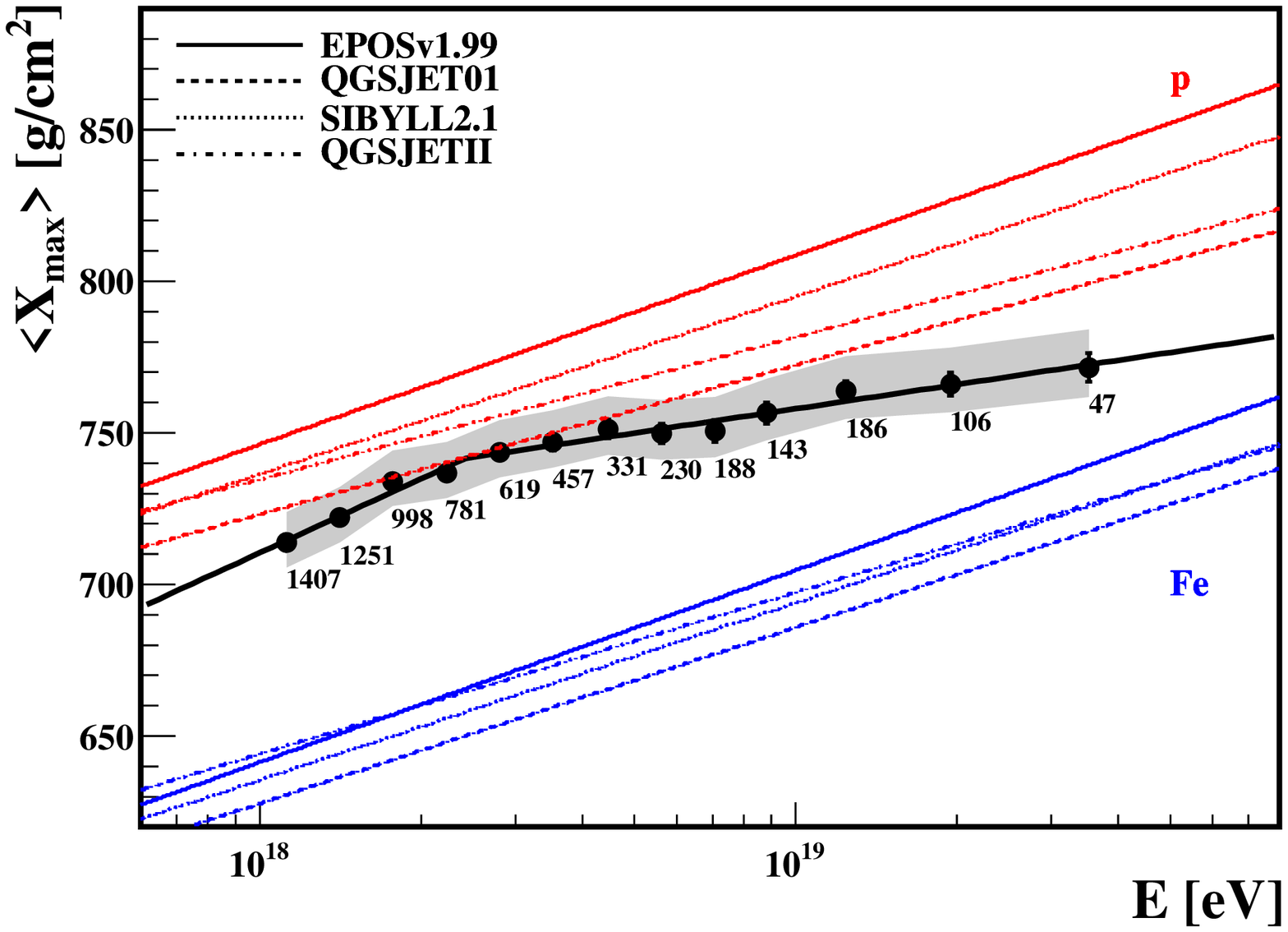,width=0.52\linewidth,clip=} &
\epsfig{file=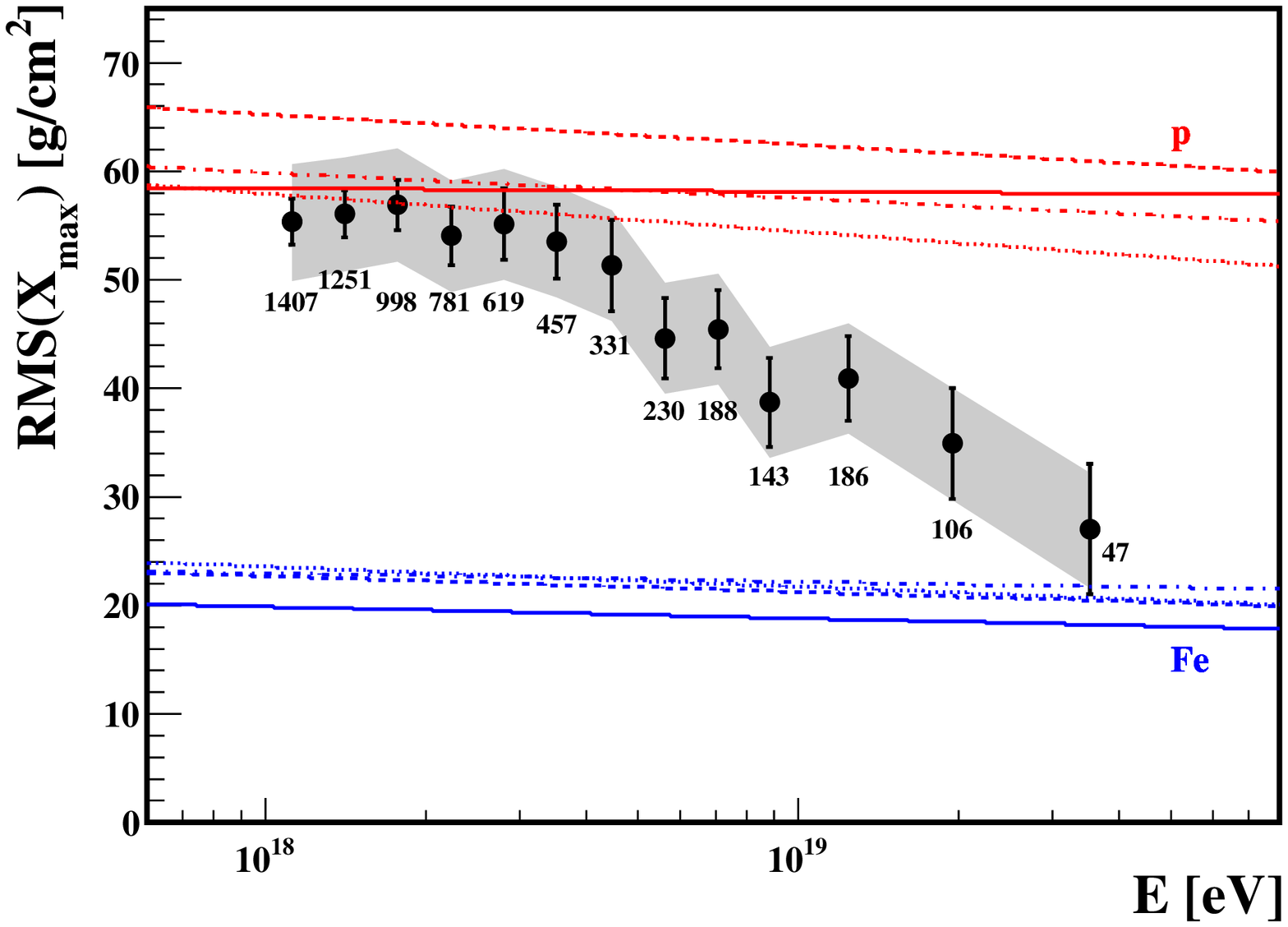,width=0.52\linewidth,clip=} 
\end{tabular}
\caption{Left: $\langle X_{\textrm {max}} \rangle$ as a function of energy. Right: $\textrm{RMS}(X_{\textrm{max}})$ as a function of the energy. In both plots data (points) are shown with the predictions for proton and iron for several hadronic interaction models. The number of events in each bin is indicated. Systematic uncertainties are indicated as a band.}
\label{f4}
\end{figure}

The measurement of the mass composition of UHECRs is essential to the solution of the problem of their origin, since the mass, and charge $Z$,  distribution can give powerful constraints on their acceleration mechanisms and propagation.
For UHECRs the main observable sensitive to composition is the $\langle X_{\textrm{max}} \rangle$, the average value of the atmospheric depth (measured in g/cm$^2$) where the shower development reaches its maximum. Proton showers are about 100 g/cm$^2$ deeper in the atmosphere than iron showers. In a similar way, the fluctuation of the values of $X_{\textrm {max}}$ around the mean depth, $\textrm{RMS}(X_{\textrm {max}})$, provides another sensitive observable: iron showers fluctuate about 40 g/cm$^2$ less than proton showers.  Those estimates are obtained from transport codes that simulate  shower development given a model for hadronic interactions. The energy evolution of the two mentioned observables  for proton and iron and several hadronic models are shown in Fig.~\ref{f4} together with the FD data \citep{comp1}, \citep{comp2}.  Both plots could be interpreted as an evolution from light to heavier composition if current hadronic interaction models adequately describe  the air shower physics.

\begin{figure}
%\centering
\hspace {-7mm}
\begin{tabular}{cc}
\epsfig{file=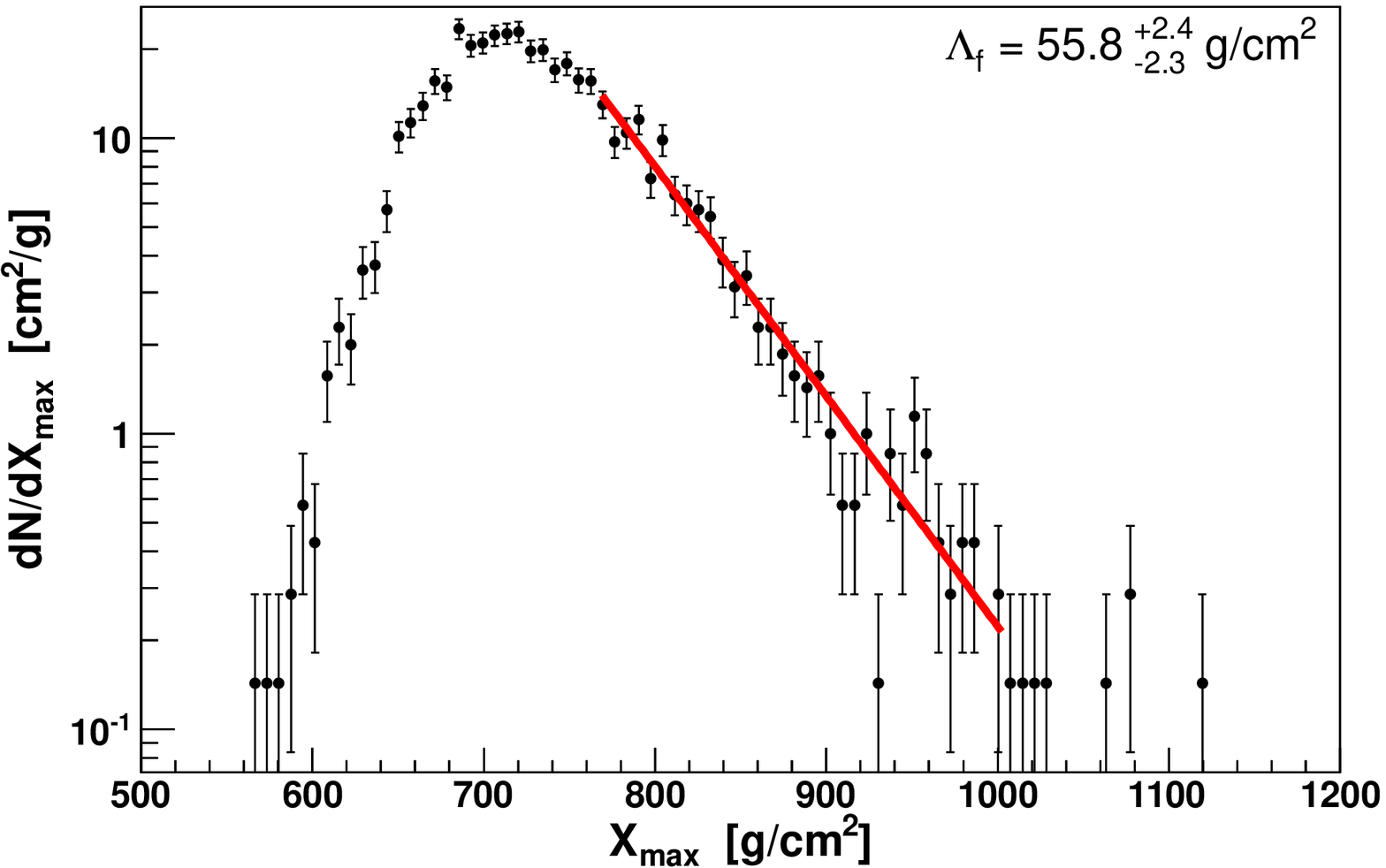,width=0.52\linewidth,clip=} &
\epsfig{file=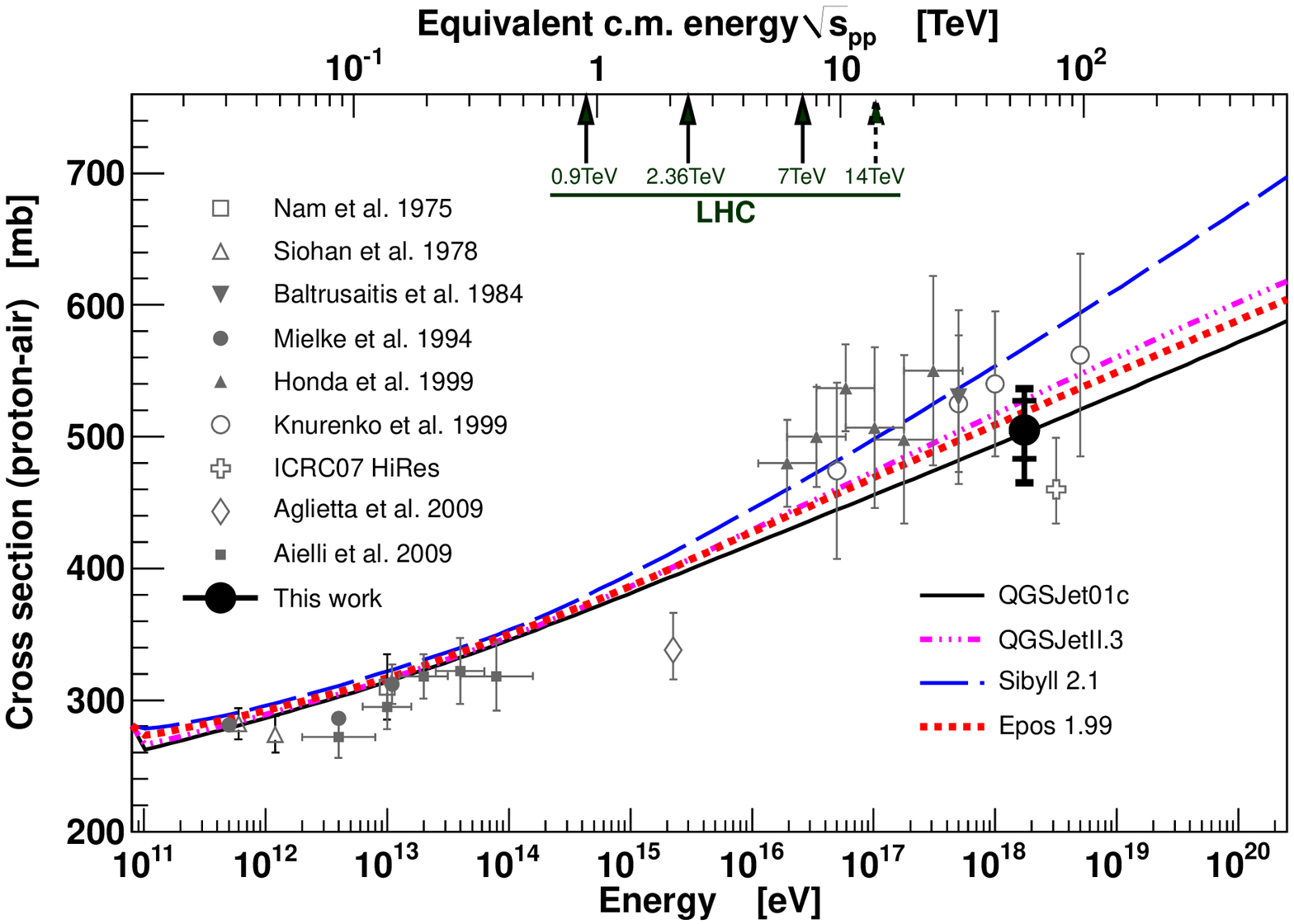,width=0.52\linewidth,clip=} 
\end{tabular}
\caption{Left: Unbinned likelihood fit of the tail of the $X_{\textrm{max}}$ distribution.  Right: Resulting $\sigma_{p-air}$ compared to other measurements and model predictions. The inner error bars are statistical only, while the outer include all systematic uncertainties for a helium fraction of 25\% and 10 mb photon systematics.}
\label{f5}
\end{figure}

An analysis of the proton-air cross-section based on the shape of the distribution of $X_{\textrm{max}}$ was presented in \citep{xsec}. That analysis is based on the fact that the tail of the $X_{max}$ distribution is sensitive to $\sigma_{p-air}$. 
The longitudinal profile of the energy deposit is reconstructed  from the light recorded by the Fluorescence detector.  With the help of data from atmospheric monitoring devices  the light collected by the telescopes is corrected for the attenuation between the shower and the detector and the longitudinal shower profile is reconstructed as a function of atmospheric depth. $X_{\textrm{max}}$ is determined by fitting the reconstructed longitudinal profile with a Gaisser-Hillas function \citep{gaisser}. The typical resolution is around 20 g/cm$^2$ above a few EeV. 
In fact, the cross section is directly related to the exponential distribution of the depth of the first interaction $X_1$ which is not accessed experimentally. The strong correlation between $X_1$ and $X_{\textrm{max}}$ makes the distribution of the latter sensitive to the proton-air cross-section and the tail of the distribution maximizes the proton content, since it is the most penetrating nucleus.   A slope obtained from a fit to the exponential tail of the $X_{max}$ distribution (see left of Fig~\ref{f5}) can be used as an estimator for $\sigma_{p-air}$ through Monte Carlo simulations: the cross-section is rescaled consistently to reproduce the value of the measurement. The lack of detailed knowledge of the mass composition at these energies turns out to be the main difficulty for this analysis, since one cannot exclude contamination by photons and helium primaries, for instance. This translates into the main contribution to the systematic uncertainty of this measurement. Only events with energy between $10^{18}$ eV and $10^{18.5}$ eV were used and the average center-of-mass energy is $\sqrt{s}=57$ TeV. If the possible presence of helium in the data sample is neglected, the measured proton-air cross-section is $\sigma_{p-air} = 505 \pm 22({\rm stat}) ^{+28}_{-36} ({\rm sys})$ mb.

\begin{figure}
%\centering
\hspace {-7mm}
\begin{tabular}{cc}
\epsfig{file=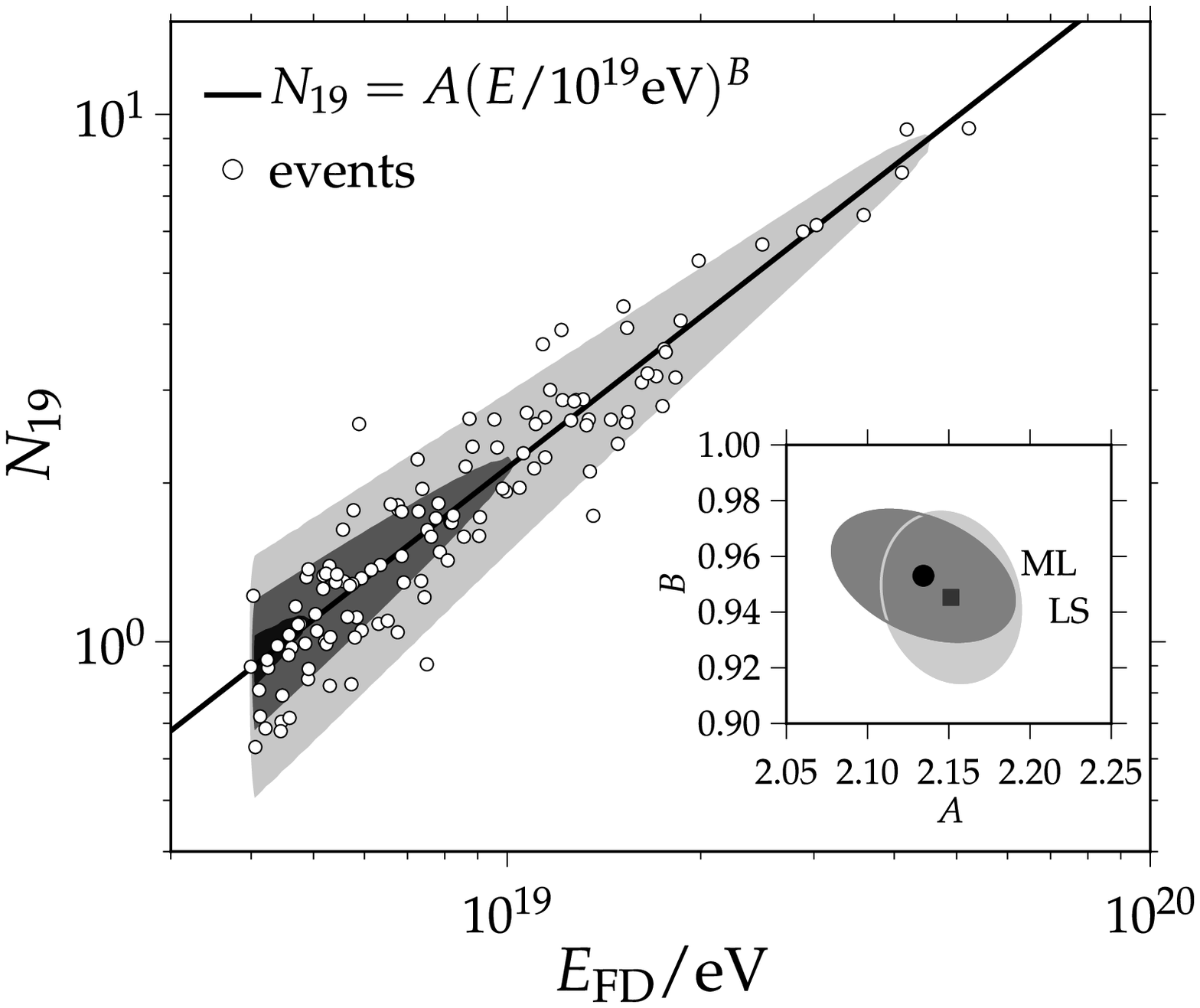,width=0.52\linewidth,clip=} &
\epsfig{file=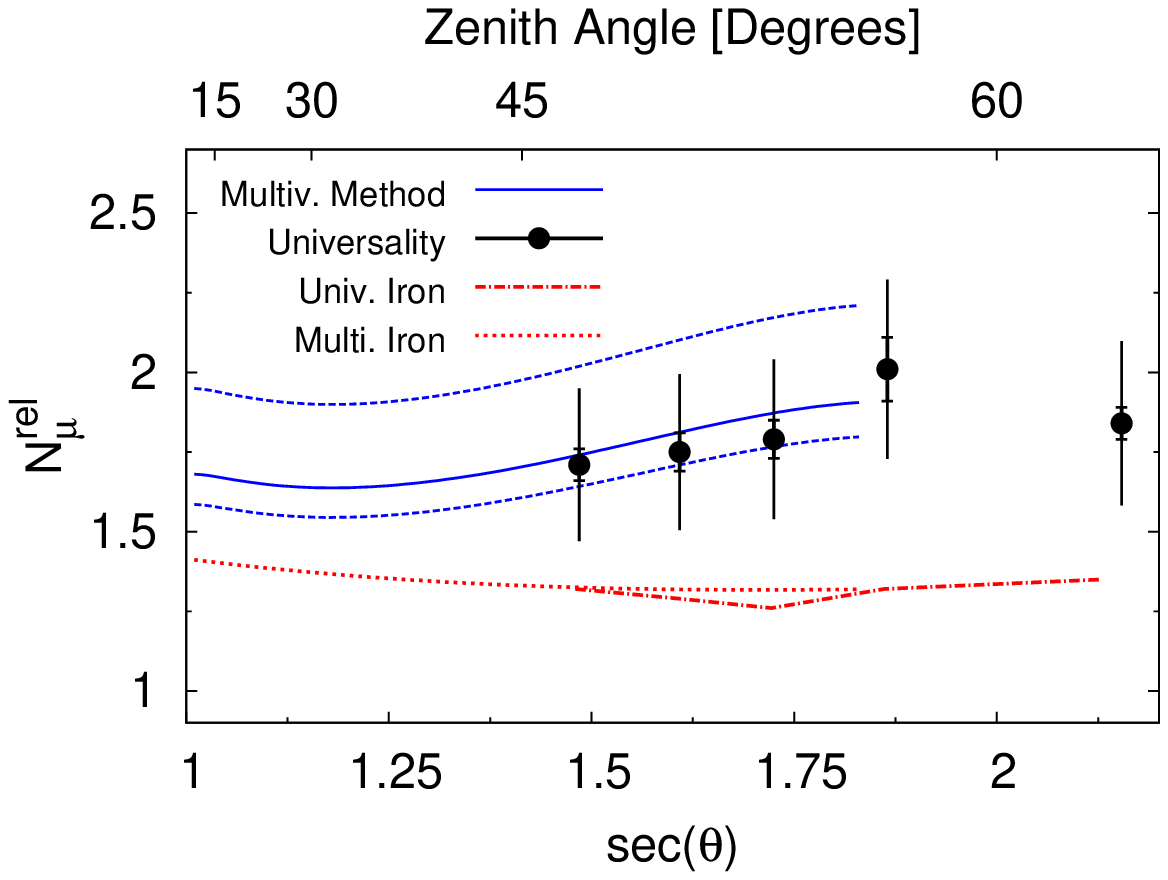,width=0.52\linewidth,clip=} 
\end{tabular}
\caption{Left:  Fit of the calibration curve $N_{19} = A(E_{FD}/10^{19} \textrm {eV})^B$  to 125 events. The contours indicate constant levels of the p.d.f. that models the observed ($E_{FD}$, $N_{19}$ and $\theta$)  integrated over $\theta$, corresponding to 10, 50, 90\% of the maximum value. The calibration constants A, B obtained with the maximum-likelihood method (ML) and a least-squares method (LS) are shown in the inset. The ellipses indicate uncertainty contours of 68\% confidence.  Right: The number of muons estimated at 1000 m in data relative to the predictions of simulations using QGSJET II with proton primaries. The results obtained using multivariate method are shown as the solid line with systematic uncertainties as the dashed lines. The results obtained using the universality of $S_{\mu}/S_{em}$ are shown
as circles with statistical and systematic uncertainties. The results when the methods are applied on a library of iron-initiated showers are shown as the dot and dash-dot lines at $10^{19}$ eV. 
 \label{f6}}
\end{figure}

Another estimator of the composition of the primary particles is the muon density on 	ground. The muon density on the ground is predicted to be 30-40\% larger for iron primaries then for proton primaries.  The value of this ratio is considered robust, as opposed to  the absolute value of the muon density that is more model dependent.

 The fraction  of the signals in the SD from the muonic component and from the electromagnetic component of the shower is measured in two ways. The most direct one to investigate the muon content of the cosmic ray showers is through the study of very inclined events \citep{n19}. In those events the dominant particles at ground are muons since the electrons and photons were absorbed by the atmosphere.
 The observable $N_{19}$ is the muon estimator that in fact is the measured shower size. For inclined showers, the energy of the primary is obtained by calibrating $N_{19}$ with the calorimetric energy $E_{FD}$ from high-quality events measured simultaneously with the fluorescence detector and the surface detector for inclined showers.  The plot on the left of Fig.~\ref{f6} shows the correlation between $N_{19}$ and $E_{FD}$.  $N_{19}$ is parameterized  \citep{n19} as a function of $E_{FD}$ with the fitting function $N_{19} = A(E_{FD}/10^{19} \textrm {eV})^B$ . For this particular choice,  $A$ is the relative number of muons with respect to the reference distribution at 10 EeV.
The slope $d\ln{(N_{19})}/d\ln{(E_{FD})}$ of the relative number of muons is given by parameter $B$, and carries information about possible changes in  $<\ln{(A)}>$.

 Using usual (not very inclined) events one  can  compare data from the SD and FD simultaneously, on a event-by-event basis, to the result of simulations, or use the time structure of the  signals in the surface detectors and an universal property of air showers to estimate the number of muons in the data \citep{muons} .  A summary of the results obtained with these two methods when applied to data is shown in the right of Fig.~\ref{f6}. 
 Simulations of air showers using QGSJET II with proton and iron primaries underestimate both the total detector signal at ground level and the number of muons in events collected at the Pierre Auger Observatory. These discrepancies could be caused by an incorrect energy assignment within the 22\% systematic uncertainty of the energy scale of the Auger Observatory and/or shortcomings in the simulation of the hadronic and muonic shower components. All results point to the fact  that simulations of proton primaries and iron primaries
underestimate the muon fraction at ground level.

\section{Conclusions}
The Pierre Auger Observatory has detected high quality events and has made key measurements of the highest-energy cosmic rays. In spite of that, many issues remain open. The spectrum suppression  is well established but it is not known if this is due to the GZK effect or if it is due to the maximum energy attained in astrophysical accelerators.  Measurements of the large scale anisotropy provides upper limits in the dipole amplitude that constrain theoretical models in the ankle region.  
The Pierre Auger Observatory data provide evidence for a correlation between arrival directions of cosmic rays above 55 EeV and the positions of AGNs with $z < 0.018$. The estimated fraction of correlating cosmic rays is currently about 33\%  compared with 21\% expected for isotropy. 

The first measurement of the cross section proton-air around  $10^{18}$ eV was presented.
 Detailed studies of hadronic interactions in the atmosphere together with a much larger sample may provide new information that will help to answer the questions raised by the UHECRs.

 %%%%%%%%%%%%%%%%%%%%%%%%%%%%%%%%%%%%%%%%%%%%%%%%%%%%%%%%%%%%%%%%%%%%%%%%%%%%%
%% Appendices
% The Appendices part is started with the command \appendix;
% appendix sections are then done as normal sections
% \appendix

\end{document}